\documentstyle[11pt,paspconf,epsfig]{article}
\def\edcomment#1{\iffalse\marginpar{\raggedright\sl#1\/}\else\relax\fi}
\reversemarginpar

\begin{document}
\title{\bf Cluster Formation and the ISM} 
\author{Ralph E. Pudritz and Jason D. Fiege}
\affil{Dept. of Physics and Astronomy, McMaster University,
Hamilton, Ontario, L8S 4M1, Canada}


\begin{abstract}
We review the physics of star formation, and its links with
the state of the ISM in galaxies.   
Current obervations indicate that the preferred mode of 
star formation is clustered.  Given that OB associations
provide the dominant energy input into the ISM, deep links
exist between the ISM and star formation.  We present a multi-scale 
discussion of star formation, 
and attempt to create an integrated vision of
these processes. 

\end{abstract}

\section{Introduction}

Star formation and the state of
the interstellar medium involve    
strongly coupled physical processes.    
This is because the primary energy inputs into the ISM
dervive from the products of massive star formation
in OB associations, namely, 
correlated supernova explosions (Type II), stellar winds, 
and photoionizing flux.  
The resulting pressure, shock waves, FUV,   
and cosmic ray content of the diffuse medium  determine  
the surface pressures and ionization of molecular clouds.
These mechanisms are the means by which
the ISM regulates the surface density of molecular clouds, as
well as the coupling of their internal magnetic fields;  
and thereby the process of star formation.

This review highlights some of the central processes in star
formation and their links to ISM physics.
We first give an overview of the basic physical 
processes in star formation  
(\S 1).  In \S 2, we review the basic physics
of molecular clouds and their cluster forming, more
massive cores.  We then highlight some current work
on the physics of filamentary molecular clouds (\S 3). 
Finally, we review some of the basic ideas of 
star formation theory with an eye to understanding
cluster formation (\S 4). 
The reader may consult recent related reviews by 
McKee {\it et. al.} (1993), McKee (1995), and Elmegreen
{\it et. al.} (1999). 

\section{Overview: From the ISM to Protostars}

\subsection{ Diffuse ISM }

The physical scales of the diffuse ISM span
the range $10^2 - 10^3 $ pc.   
The diffuse medium consists of cooler clouds that are in
pressure balance with a hot surrounding medium, following
the ideas of Field (1965), and McKee and 
Ostriker (1977, henceforth MO).  The existence of a multiphase medium
hinges upon the structure of the cooling function
of atomic gas. 
The diffuse ISM has 
three phases in its inventory (see review; McKee 1995):
 
$\bullet$  {\it cold neutral medium} of clouds (CNM)
with a temperature $T \simeq 50^o$ K and density $n \simeq 40 cm^{-3}$;
 
$\bullet$ {\it warm neutral medium} (WNM) of temperature $T \le 8,000^o$ K
which surrounds the CNM and pervades much of space and,
a {\it warm ionized medium} (WIM) at $T \simeq 8,000^o$ K and
$n \simeq 0.26$ cm$^{-3}$ which
consists of highly ionized warm hydrogen;
 
$\bullet$
 {\it hot ionized medium} (HIM) with $T \simeq 5 \times 10^5$ K and
$n \simeq 10^{-2.5} cm^{-3}$.

The ISM contains other important components however. 
Of greater significance, energetically, is the pressure associated
with the magnetic field that pervades the ISM.  It has
a strength of 
$B \simeq {\rm several \quad \mu G}$.  Cosmic rays have comparable
energy densities and are thought to be accelerated in 
supernova remnants.  Thus, the bulk of the cosmic ray
component of the ISM 
can be viewed as a consequence of star formation 
(see Duric, this volume).   
Finally, the ISM has  
non-thermal gas motions whose energy density is comparable
to the magnetic and cosmic ray contributions as well.
The fact that the energy density in the non-thermal 
component dominates the thermal
one is reminiscent of the physics of molecular gas, which
we address below.

The total pressure in the ISM can be ascertained 
by applying the condition of hydrostatic balance: 
the pressure in the galactic plane must be sufficient to 
support the weight of a hot galactic corona, detected through
X-ray emission.  The result (see Boulares and Cox, 1990) is that  
$\bar P_{ISM} \simeq 
(3.9 \pm 0.6 ) \times 10^{-12} $ dyne cm$^{-2}$.
The constituent pressures that add up to this value are,
in units of $10^{-12}$ dyne cm$^{-2}$ 
(from Boulares and Cox):
 
$ \bullet$ thermal pressure, $\simeq 0.3$
 
$\bullet$ cosmic rays, $\simeq 1.0$
 
$\bullet$ magnetic fields, $\simeq 1.0$ (field strengths locally 2-3
$\mu$G)
 
$ \bullet$ kinetic pressure, $\simeq 1.8$.  

\noindent 
The thermal energy contributes little to the
overall budget because   
the gas cools so efficiently, the  
primary coolant being the CII line.  
Thus the total energy budget for the diffuse ISM 
may be written as,


\begin{equation}
E _{therm} << E_{non therm} \simeq E_{CR} \simeq E_{mag} >> E_{grav}
\end{equation}

The MO theory predicts that the hot phase of the ISM is heated
by supernova explosions.  The SN are 
assumed to occur at random throughout the Galactic disk.
Uncorrelated Type I supernovae as well as  
a distributed Type II population are the drivers
of ISM energetics in this picture.  The 
HI survey of the Milky Way
by Colomb, Poppel, and Heiles (1980)
revealed, however, that vast atomic clouds are organized
in filamentary structures. 
Here and there, large
cavities and shells are obviously discerned.  
Heiles (1984) catalogued a large number of large HI shells in
the ISM which he suggested are produced by the combined
energy input from winds and supernova explosions
that accompany OB associations. 
Note that OB stars formed in associations account for
90 \% of the supernovae in the galaxy (van den Bergh and
Tammann 1991).    

Theoretical models of
the energy input from massive stars in the HI of
the galactic disk show that
the correlated massive stellar 
winds and supernovae in OB associations create large superbubbles
in the galactic HI (Bruhweiler {\it et. al.} 1980).   
The swept up
gas forms dense, compressed supershells.
Filaments could
be produced as these shells cool
and undergo thermal instabilities.
Sufficiently intense wind and supernova activity
can also lead to break-out into the galactic halo.
The plumes of hot gas flowing vertically out of the disk
are akin to "chimneys" located above the OB associations
(eg. Norman and Ikeuchi 1989). 
A beautiful example of this process in action is
the CGPS 21 cm map of the chimney in the W3 region
(Normandeau {\it et. al.} 1996). 

Large scale interstellar magnetic fields can strongly
ffect the evolution
of superbubbles.  Recent numerical calculations by
Tomisaka (1998) as an example, show that blow-out depends upon
the structure of the magnetic field in the ISM. 
For a uniform field parallel to the galactic
plane, expansion perpendicular to the plane is 
suppressed and a strongly magnetized shell
is swept up.  On the other hand, if the field strength
scales with the gas density as $B \propto \rho^{1/2}$,
superbubble expansion and blow-out occur more
readily, the results being similar  
to the unmagnetized case.   
These calculations also show that the field is stretched
in the walls of the expanding shell and appears to be   
well ordered.  

We now turn to the question of how the ISM 
affects the physical properties of molecular clouds.
There are at least two important influences: 

$ \bullet$  ISM Pressure  $\rightarrow$ Molecular Cloud Properties

$ \bullet$  Cosmic Ray production $\rightarrow$ Molecular Cloud Heating/Ionization

\noindent
The ISM pressure 
upon the surfaces of molecular clouds truncates
them and in so doing,  determines their size. 
Thus, the ISM pressure essentially determines the surface density
of a cloud.  The second point is that
the ionization  of molecular clouds is determined by two
types of radiation; FUV photons and
high energy cosmic rays.  The cosmic rays penetrate into
the heart of high column density clumps in the cloud
and partially ionize the gas, while FUV photoionization 
dominates in gas with visual extinctions $A_V \le 4$ mag
(McKee 1989). 
By contributing to the ionization rate in a cloud in this
way, the
ISM activity
helps to control the star formation rate.

\subsection{Molecular Clouds} 

Stars form in self-gravitating, 
molecular clouds.   These are highly inhomogenous
structures that are filamentary (eg. Johnstone and Bally, 1999), and have  
a rich sub-structure.
Molecular clouds are host to a spectrum of dense, 
high pressure sub-regions
known as molecular cloud clumps and smaller cores.  These  
are the actual sites of star formation
within molecular clouds.   

Most of the self-gravitating gas in the Milky
Way is gathered in the distribution of Giant Molecular
Clouds (GMCs).  Surveys show (see eg. Scoville and Sanders, 1987)
that GMCs range in mass from $10^5 - 10^6 M_{\odot}$
although smaller molecular clouds may be as low as several $10^2 M_{\odot}$.
GMCs have a mass spectrum, $dN/dM_{GMC} \propto M_{GMC}^{-1.6}$. 
The range in physical scales is about    
$10 - 100$ pc, with the median cloud having a mass of
$3.3 \times 10^5 M_{\odot}$ and a median radius 
of 20 pc.  The median cloud density is then
180 cm$^{-3}$, and column density
$1.4 \times 10^{22}$ cm$^{-2}$ or equivalently,  $260 
M_{\odot} $ pc$^{-2}$ (see Harris and Pudritz 1994).  
The pressures within molecular clouds are much
higher than the ISM pressure surrounding them. Typically 
$P_{GMC}/P_{ISM} \simeq 20-30$.  Thus, molecular clouds
are not another phase of the interstellar medium because self-gravity,
not pressure, dominates their internal physics.     
Direct Zeeman measurements of molecular clouds also
show that they are pervaded by strong magnetic fields 
$B \simeq 30 \mu$G. 
The corresponding magnetic pressure is 
comparable with the gravitational energy density of the
a cloud.   
 
The line-profiles of molecular clouds are dominated by non-thermal motions.
Larson (1981) was able to establish two important empirical
cloud properties.  If $r$ represents the physical scale 
of a CO map, and one measures the average column density $\Sigma$ 
of gas within this scale, Larson found that
$\Sigma \propto \rho r \simeq const$.  Moreover, if
$\sigma$ is the velocity dispersion of the gas on this
scale, he also determined that $\sigma \propto r^{1/2}$.
The physical explanation of these relations came with
the work of 
Chieze (1987) and  Elmegreen (1989).    
The theory of self-gravitating spheres that are embedded within
an external medium of pressure $P_s$ was worked out
in classic papers by Ebert (1955), and Bonner (1956).
They showed that an isothermal cloud of mass $M$ and
average velocity dispersion $\sigma_{ave}$ has a critically
stable state, described by 

\begin{equation} 
 M_{crit} = 1.18 {\sigma_{ave}^4 \over (G^3 P_s)^{1/2}}
\end{equation}
\begin{equation}
 \Sigma_{crit} = 1.60 (P_s/ G)^{1/2}\ 
\end{equation}
We note that these relations may also be determined 
from the virial theorem (eg. McCrae 1957). 
They give a natural explanation for Larson's
empirical laws.  The surface density of a molecular cloud,  
so interpreted, depends only upon 
the ISM pressure $\Sigma \propto P^{1/2}_{ISM}$.  It is the   
relative constancy of the ISM surface pressure
over significant portions of the disk then, 
that accounts for the near constancy of molecular cloud surface
densities. 
Similarly, by noting that $ M = \pi r^2 \Sigma$, one sees 
that the first equation 
yields Larson's size-line width relation.  Exactly the same type
of scalings, as well as coefficients of the same order of magnitude
pertain to the description of a {\bf filamentary cloud},
as Fiege and Pudritz (1999, FP) show.

The energetics of a molecular cloud may be summarized as
follows;
\begin{equation}
E_{therm} << E_{nontherm} \simeq E_{mag} \ge (1/2) E_{grav}
\end{equation}
where the last inequality follows from the fact that 
ordered fields, and non-thermal (MHD) motions contribute
equally to GMC cloud support (eg. review McKee {\it et. al.}, 1993).
The explanation of the insignificant contribution of thermal
support to this balance is again related (as in the ISM)
to the efficiency of cooling;  
the predominant coolant in molecular gas now being the CO rotational lines
$J = 1-0 $ etc.  By balancing the heating rate of 
molecular clouds due to cosmic ray bombardment,
with the molecular cooling rates 
one can  show that clouds are
maintained at a temperature of     
$ 10-20^o K$ over a wide range of densities
$10^2 - 10^4 $ cm$^{-3}$ (eg. Goldsmith and Langer 1978).  
It turns out that the column density at which 
HI clouds attain sufficiently high column density,
( several $ 10^{21}$ cm$^{-3}$) to become self-gravitating, is  
also the column at which self-shielding from 
the galactic FUV field takes place allowing the 
gas to become molecular.  

The onset of molecular
cooling in sufficiently high-column gas drops the cloud to 
low temperatures where it is thermostatically
controlled.   Molecular clouds also have low rotational
energies compared with gravity.  Thus, 
it is
left to magnetic pressure as well as 
the non-thermal motions (MHD waves or turbulence)
to provide support of the cloud against gravitational
collapse.  Molecular clouds are exotic structures
in which magnetism plays out a slow, and ultimately losing
battle against self-gravity.    

The literature contains  four basic
{\bf mechanisms for cloud formation} that may all be active in 
the galaxy:

$\bullet$ Cloud-cloud agglomeration in spiral waves (eg. Kwan \& Valdes 1983)

$\bullet$ Supershell fragmentation (eg. McCray and Kafatos 1987)

$\bullet$ Large-scale gravitational instability:
{\bf $Q$ criterion} (Kennicutt, 1989)

$$ Q = { c_s \kappa \over \pi G \Sigma } < 1  \quad {\rm gas} $$

$\bullet$ Parker Instability (eg. Blitz \& Shu 1980, Elmegreen 1982)

\noindent
The evidence for the first comes from CO observations
of spiral galaxies.  One invariably finds that GMCs inhabit
the spiral arms of galaxies (eg. Rand 1993).  The idea is that 
the passage of an arm leads to focussing of orbits
of more diffuse clouds.  These collide and agglomerate
in the potential well of the arm.  The form of the
power law mass spectrum of clouds, given above, is
in good support of this mechanism as Kwan and Valdes
showed.  The second mechanism also appears to 
operate; within the walls of supershells.   
The fragmentation mass is of order $ 5 \times 10^4 M_{\odot}$
by the time that $3 \times 10^7$ years have elapsed
(see McCray and Kafatos, 1987).     
The third mechanism, that of gravitational instability in 
a self-gravitating gas layer, employs the 
well-known Toomre 'Q' criterion where $c_s$ is the effective
sound speed of the medium and $\kappa$ is the epicyclic
frequency.  Kennicutt (1989) showed that significant star
formation in spiral galaxies only occurs in gas whose
column density exceeds the limit given by the Toomre criterion, 
and is related, therefore, to gravitational instability 
of the gas.  
Finally, the Parker instability
arises from the intrinsic buoyancy that a magnetic field,
oriented parallel to the galactic plane, has
within a self-gravitating gas layer.  As the field rises and bubbles
out of a gas layer, the gas flows back
to the disk along magnetic field lines and gathers in 
magnetic "valleys".  The predicted scale length of this
instability is about 1 kpc along spiral arms
(eg. Elmegreen 1982), which
is precisely the kind of distance one sees between
GMCs and giant HII regions in galaxies. 
The complete picture of GMC formation
probably involves all of these processes.  
Gas clouds agglomerate   
within spiral arms and finally reach the critical
column density at which gravitational instability,
and magnetic buoyancy become important.     
The subsequent formation of OB associations and supershells 
leads to further gas compression, and also the possible
formation of chimneys that channels some of the energy
release into the galactic halo.

\subsection{Molecular Cloud Clumps: Cluster Formation}

The clump mass spectrum within molecular clouds 
is, interestingly enough, similar to the GMC mass
spectrum.  Thus, 
Blitz (review, 1993) reported that  
the spectrum  $dN/dM \propto M^{-1.6}$, with a 10 \% 
error in the index, fit the data for a number of 
different clouds rather well. 
The masses of such clumps range from 
$ 1 - 1000 M_{\odot}$,   
so that one sees that the median clump mass is only
a thousandth of the median GMC cloud mass.  
The fact that only a few percent of the 
mass of a molecular cloud is tied up in these star
forming clumps is, empirically, the reason why star formation is 
a rather inefficient process.

The overall energetics of clumps are more dominated by 
gravity than the diffuse inter-clump regions of the 
cloud.  
These clumps have strong fields and their
internal kinematics are dominated by 
non-thermal motions (eg. Casselli and Myers, 1995).  
Maps of the magnetic structures associated
with clumps are now
becoming available through the technique of submillimetre 
polarimetry.  Schleuning's (1998) map of a clump in 
the Orion cloud as an example, shows that there is a well ordered,
hour-glass shaped magnetic field structure present
on these scales. 
Not all clumps are expected to lie near the
critical threshold for gravitational collapse.  
As an example, low mass clumps in the logatropic model for
cloud cores developed by McLaughlin and Pudritz (1996)
are pressure dominated structures.  In that model, a critical
magnetized clump has a mass of $\simeq 250 M_{\odot}$. 

Infrared camera observations of clumps in molecular clouds
such as Orion reveal that the most massive clumps
are forming star clusters.   The most massive clump in 
the Orion cloud as an example, has a mass of $\simeq 500
M_{\odot}$ and has a star formation efficiency approaching
40 $\%$ (E. Lada 1992; also reviews by Zinnecker {\it et. al.} 
1993, Elmegreen {\it et. al.} 1999).  
Stars typically form as a member of a group or
cluster, and not in isolation.     
This may be explained by noting
that clump mass spectra
are rather different than the Salpeter IMF for the stars.
Beyond $ 0.3 M_{\odot}$ or so, the IMF has a
much steeper power-law than that describing molecular
clumps; $dN_*/dM_* \propto M_*^{-2.35}$. 
By weighting the clump mass spectrum with the mass, and integrating
over mass, one finds that the total mass of the clumps scales as 
$ M_{clump, tot} \propto M_{max}^{0.4}$.  Thus, most of the mass
resides in the massive end of the spectrum of clumps.
For stars however, this same procedure shows that it is 
the {\it lower mass end } of the IMF that contains most 
of the stellar mass.  Therefore, cluster formation   
{\it must be} the preferred mode of 
star formation (Patel and Pudritz 1994).

How do the clumps form?   
Processes that have been proposed include;

$\bullet$  Clump-clump agglomeration (Carlberg and Pudritz 1990, McLaughlin
and Pudritz 1996)  

$\bullet$  Structures in turbulent flow (see Pouquet's contribution).  

\noindent
The fact that clumps and 
their parental clouds share similar mass spectral
forms suggests a common formation mechanism. 
Thus clumps could be built up by the
same type of agglomerative and gravitational instability 
process that 
produced molecular clouds.  Turbulence
has also been proposed as 
the origin of a wide range of structure and substructure
(Elmegreen and Efremov 1997).     

The finite amplitude Alfv\'en waves within
molecular clouds and their clumps contribute
not only to the support of the gas, but also
to the process of clump and structure formation
within them.  Dewar (1970) showed that
a collection of Alfv\'en waves exerts a nearly
isotroptic gas pressure that can support
gas in a direction parallel to the magnetic field
lines.  Arons and Max (1975) later championed
the idea that such waves might account for
non-thermal motions in clouds.    
Alfv\'en waves of finite amplitude are different
than linear waves however,  in that they produce density
fluctuations.  
Carlberg and Pudritz (1990)  
suggested that such non-linear waves could
produce the small density fluctuations   
within clouds that subsequently agglomerate to produce the
observed clump mass spectrum.  

This behaviour of non-linear Alfv\'en waves 
is akin to pressure
(magneto-acoustic) waves, which damp out quickly. 
Numerical simulations of the behaviour of such
waves find that   
they damp very quickly;  
with $\tau_{damp} \le 2 t_{ff}$ 
(see 
Vasquez-Semadeni, Passot and Pouquet 1995, Maclow et al 1998, Ostriker
et al 1998).
Thus, unless MHD waves and turbulence are replenished very quickly,
the turbulent support of clumps would quickly vanish.
It has often been suggested that bipolar outflows, examined below, can solve
this problem.  One caveat here is that turbulence
is also observed in starless clumps and cores. 

\subsection{Molecular Cloud Cores: Individual Star Formation}

On scales of  
0.01-0.1 pc one encounters dense cores
($n_{core} \ge 10^4$  cm$^{-3}$) of several solar masses that may be 
associated with the
formation of individual protostars (Benson and Myers, 1989).  
Their internal
velocity dispersions are less dominated by non-thermal
motion 
(see Casselli and Myers 1995).  

We have arrived at perhaps the most intensively studied level of 
star formation. 
Individual star formation is an incredibly rich process
involving the simultaneous presence of;  

$\bullet$  Accretion disks

$\bullet$   Bipolar outflows and jets

$\bullet$  Gravitational collapse

\noindent
Jet activity and accretion disks are strongly correlated
because whenever one sees and outflow, there is 
good evidence that an accretion disk is also present.  
Bipolar molecular outflows are the most obvious and ubiquitous sign-post of star
formation. There are now more than 200 molecular outflows known 
(eg. reviews; Bachiller 1996,  
Padman, Bence and Richer 1997).
They persist for at least several
$10^5 $ years, which is a good fraction of the pre-main-sequence
evolution timescale. 
The outflows consist of material in the natal molecular cloud
core of a protostar that is swept up and put into motion by a faster, underlying
jet from the central source. Outflows are associated
with stars of all masses;  from the common
low-mass T-Tauri stars, to high mass stars in the process of  
formation.
A model for the underlying jets that appears to fit all of the observed
facts predicts that jets are centrifugally accelerated winds
driven from the surfaces of magnetized protostellar accretion
disks (eg. review, K\"onigl and Pudritz 1999).   
Another suggestion is that jets originate at the interface between
the magnetosphere surrounding an active young protostar, and the
surrounding disk (eg. review Shu {\it et. al.} 1999). 

{\bf How do cores form?}  Several ideas have been suggested in 
the literature, including:

$\bullet$ Ambipolar diffusion (eg. Mouschovias 1993)

$\bullet$ MHD wave damping ( Langer 1978,
Mouschovias 1987,  Pudritz 1990)

\noindent
The first mechanism arises from the fact that 
the magnetic field within
a partially ionized gas  
(typically one ion in $10^7$ neutrals in 
a molecular cloud) is fairly diffusive.  
A field will slowly slip out of 
overdense regions because 
in a magnetic field gradient,
the Lorentz force pushes outwards on the ions
to which the field is directly coupled.  Collisions
of the slowly outward moving ions 
with the neutrals communicate the magnetic
force to the overwhelmingly neutral gas. 
This magnetic pressure support cannot be
maintained forever because the ions are gradually 
pushed out of the dense
region, dragging the field with them.  The  
neutral density, on the other hand, slowly increases.
The drift time scale of the field $v_D$ is
thus proportional to the gravitational acceleration
$g$ in the clump times the time-scale for a neutral
to collide with any ion, $\nu_{ni}^{-1}$.  
The ambipolar diffusion time-scale is therefore
(eg. McKee et al 1993), 
$ \tau_D = (R / v_D) = (R / g \nu_{ni}^{-1}) \propto (\zeta /
\rho)^{1/2}
$
where $\zeta$ is the ionization rate per molecule.  
Compare this with the free-fall time;
$ t_{ff} = (3 \pi / 32 G \rho)^{1/2} = 1.4 \times 10^6 (n / 10^3
cm^{-3})^{1/2} \quad {\rm yr}.  $ 
Both these time scales have the same power-law
dependence upon the  cloud density.  Thus, their   
ratio,  
$ \tau_D / t_{ff} \simeq 10$ for standard
ionizing flux rates.  This result is {\it independent of the
cloud density}.  We see that 
magnetic pressure can stave off gravitational collapse
for a significant number of dynamical time-scales, but that
the battle is finally lost.  The time-scale required
to form a core from the background molecular cloud
(of density $10^2$ cm$^{-3}$) has been calculated by
taking into account a host of complicating factors such
as the effect of grains on the ionization state of the
gas.  The results (Ciolek and Mouschovias 1995) suggest
that it would take nearly 20 million years to form
a core.  This may be too long in comparison with 
the lifetime of a molecular cloud, because it suggests
that most cores should  
appear to be starless.

The process of ambipolar diffusion and 
the gradual loss of magnetic support leading to 
gravitational collapse is the 
{\bf current paradigm for star formation}
in a molecular core 
(eg. review Shu {\it et al} 1987).  By taking into
account the  
angular momentum of the parent core,
the collapse gives rise to the formation of 
a magnetized disk.  While attractive, 
one may well wonder whether or not this
vision of individual star formation is able to account for 
the formation of star clusters?  
Of particular concern is the question of what determines
the mass of a star in this scenario.
We address
this concern in \S 4.

The second mechanism for core formation begins  
with the fact that MHD waves in a partially 
ionized medium are damped on a small enough scale.
The physical idea is that a wave in the combined ion-neutral
fluid can only be sustained if a neutral particle collides with
any ion within the period of an Alfv\'en wave.  
It is easily shown
that on scales of approximately the size of cores,
this condition breaks down and that waves
must damp (eg. Mouschovias 1987,  Pudritz, 1990).  
Without this wave support, gas may be more
prone to increase its density, and to eventually 
collapse.

\section{Molecular Cloud Structure: Topology and Helical Fields}

Let us now consider models for filamentary clouds, namely 
upon magnetized, pressure-truncated 
cylinders.   
An analytic solution for
the radial structure of a self-gravitating, isothermal filament
(without pressure truncation) was found by Ostriker (1964);

\begin{equation}
\rho={\rho_c \over (1+(r^2 /8 r_0^2))^2}.
\end{equation} 

\noindent
The point to note about this solution is
the steep fall-off of the gas density at larger scales;
$\rho \propto r^{-4}$.  Such a steep density gradient
has never been seen in any study of molecular clouds. 
Molecular cloud core density profiles are never steeper
than $r^{-2}$.  Recent infrared-studies of the
molecular filaments 
(eg. Lada {\it et. al.}, 1998) find radial
profiles that fall off as $r^{-2}$.  These results
raise the basic question; what is a self-consistent
model for filamentary clouds?

The MHD models in the literature posit predominantly 
poloidal magnetic fields (eg. Nagasawa 1987,
Gehman {\it et. al.} 1996).  Such a field contributes
its pressure to the  
support against gravitational collapse.  What is often
ignored in such treatments is the effect of external
pressure, as well as the possibility that filaments
have a toroidal component of the field as well.  
The possibility that filaments are associated
with helical magnetic fields was raised in the observations
by Heiles (1987), and discussed by Bally (1987).

How would helical fields arise?
All that is required  
is to twist one end of a filament containing
a poloidal field, with
respect to the other end.  This could easily occur
in the interstellar medium. 
We (Fiege and Pudritz 1999, FP; see Fiege and Pudritz, this
volume) have 
developed a rather general model 
in which we idealize molecular filaments as infinitely long cylinders 
of self-gravitating gas, that are truncated 
at a finite radius by the pressure $P_s$ of the external medium.
The general
internal field consists of both  
poloidal $B_z$ and toroidal $B_{\phi}$ field
components.  We also include an internal,  
non-thermal gas pressure.   
One then  
applies the tensor virial theorem (FP)
to derive a new form of the scalar virial theorem 
that is appropriate to the radial equilibrium of 
an infinite cylinder.

In the absence of an ordered magnetic field;
the virial theorem for pressure truncated cylinders 
takes the simple form;
\begin{equation}
{m \over m_{vir}} = 1 - { P_s \over <P>}, 
\end{equation}
\noindent
where $m$ is the mass per unit length of a self-gravitating
cylinder whose average internal pressure is $<P>$, and 
$m_{vir}=2 <\sigma^2> /  G $ 
is a critical mass per unit length (see eg. McCrae 1957).
This solution demonstrates that  
the stability of
cylinders is different than for spheres. 
In the latter, a cloud of fixed mass, when squeezed
with ever higher external pressure, reaches a critical
radius below which the solution is unstable and 
spherical collapse ensues.  For the cylinder on 
the other hand, as long as its mass per unit length  
is small enough that, $m < m_{vir}$, it doesn't matter how
much you squeeze; the filament will remain stable
against radial collapse for any external pressure. 
Solutions to this virial equation depend upon observationally
measurable quantities such as the ratio,  
$P_s/<P> $ and  $m/m_{vir}$.  We compare 
our models with the available data by plotting these 
two  
virial parameters for a number of observed filaments.

Let us now turn to magnetic field effects (FP). 
A poloidal field adds magnetic pressure support to the
filament, so that its critical mass per unit length
increases.  For a purely toroidal field however,
the opposite trend occurs because the hoop stress associated
with this component squeezes the filament. A toroidal
field also produces a much shallower density gradient.
By including these two field components in the virial
theorem, one can draw curves through the two dimensional
observational plane discussed above. 
We plot, in Figure 1, the predictions of the magnetic version of the
virial theorem.  
The observer is not required to have any field strengths
available, but must have enough data on hand to ascribe
to any filament, the two ratios (involving the pressure and
mass per unit length) discussed above.  
Each curve
represents the solution for a different relative contribution
of poloidal and toroidal field components to the 
virial equation, as measured
by a magnetic virial parameter.   For the limited
amount of data available, it appears that
filaments lie in a magnetic field regime
wherein the pinch of the toroidal field
dominates the support of the poloidal field component
(see FP).  Helical fields, it appears, may explain the data. 
A very important vindication of this result is that 
{\it the radial density profiles for filaments with helical fields
follow $\rho(r) \propto r^{-2} $}, or shallower, in agreement
with the current data. 

\begin{figure}
\hspace{\fill}
\epsfig{file=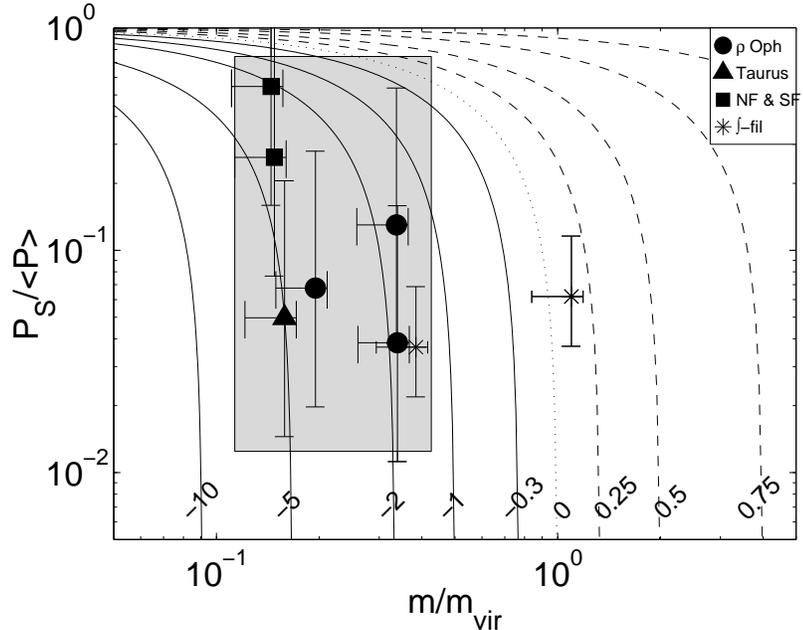,width=0.8\linewidth}
\hspace{\fill}
\caption{Helical field models are compared with the 
observed properties of real
filaments.
Curves are shown for various values of a magnetic virial parameter 
(see FP). 
Positive
values, corresponding
to the dashed curves,
indicate that the poloidal field is dominant, 
while negative values, corresponding 
to solid curves,
indicate that the
toroidal field is dominant.  The dotted line represents all solutions that are n
eutrally affected
by the helical field (including the unmagnetized solution).  
The integral-shaped 
filament appears twice, because we have used
two independent data sets in our analysis. Adapted from FP}
\end{figure}
\section{The Clustered Mode of Star Formation}

We have seen that OB associations provide 
the dominant input into heating and sculpting the properties
of the ISM.   It also appears that star formation
is dominated by the process of
cluster formation.  Thus, from both the point of 
view of star formation, and ISM theory, it
is imperative
that we understand how stellar clusters are formed.
The crucial question is, what processes determine
the mass of stars  
and hence, the form of the IMF?  

As is well known, there are two
physical pictures of invidual star formation that              
have dominated the discussion in the
last decades (eg. review; Pudritz {et. al.} 1996).  
The first view,  based
on the gravitational stability of B-E spheres,
posits that the mass of a star is essentially a    
{\it Jeans mass}.  
At a temperature of $10^o$ K, and for reasonable core 
pressures, the critical B-E mass is indeed around
$1 M_{\odot}$ (eg. Larson 1992).  
The basic problem with this approach, 
however, is that
in a highly inhomogenous medium the Jeans mass becomes
an ambiguous quantity. 
In order to explain the  
IMF therefore, one would have to posit that it is
"laid-in" in the mass spectrum of clumps.  We have noted
that this is unlikely for larger mass scales.  However,
recent submm observations by Motte {\it et. al.} (1998)
of clumps in 
the $\rho$ Oph cloud claim to have found a core spectrum
much like the Salpeter IMF.  

The second basic model for star formation
is the idea that it is not the mass, but rather the
{\it mass accretion rate} $\dot M$ that is determined
by a molecular cloud.  The self-similar collapse
picture, (eg. Shu 1977), require
that some process {\it turn-off} the infall.  The
observed outflows have been hypothesized to play this
role.  Note however, that jets in Class 0 objects are
highly collimated and will not intercept most of the
infalling envelope.  In addition, there is a strong
positive relation between disk accretion, and jet outflow
which seems to indicate that plenty of matter can collapse
onto a disk, and accrete through it onto the central
protostar.   

Both of the previous pictures appear to be inadequate
to provide a theory 
for the IMF.  A promising new approach
however, posits that stellar mass is aquired through
a process of competitive accretion for the gas
that is available in a clump.  
In the specific numerical simulation of Bonnell {\it et. al.} (1997) 
as an example,
the accretion rate is modelled 
as a {\bf Bondi-Hoyle} process onto distributed set
of initial objects of mass $M_o$.  The stellar velocity is
$v_{\infty}$, and
the gas density of the clump is $\rho_{\infty}$.   
The simulations showed that seed objects in the higher,
central density region became more massive than the 
outliers.  
This can be understood analytically since
the time dependence of accretion
at the Bondi accretion rate, given by  
$
\dot M_{BH} = \gamma M^2
$
where
$
\gamma \equiv 4 \pi \rho_{\infty} {G^2 / (v_{\infty}^2 + c_s^2)^{3/2}}
$
is readily found.  It is easy to show that   
the accretion time scale in this picture 
is
$ t_{accr} =
                  ( 1 / \gamma M_o) \propto (1 / \rho_{\infty} M_o) 
$
Shorter accretion times occur in higher density regions
of a clump, which is where one might expect more massive
stars to be formed.  
These results appear to  
strongly dependent on mass of input objects which is not
determined by the theory.  Nevertheless, this type of
approach holds considerable promise for a theory of
cluster formation. 

\acknowledgements

We are grateful to the organizers of this
meeting for the invitation to participate
in this most stimulating work shop.
We thank Chris McKee and Dean McLaughlin for
interesting discussions about these topics.
REP's research is supported through a grant
from the Natural Sciences and Engineering 
Research Council of Canada.

\end{document}